\documentclass{ws-procs9x6}

\begin{document}

\title{Measurements of $\alpha_s$ and parton distribution functions using HERA jet data}

\author{A. M. Cooper-Sarkar}

\address{Oxford University Physics Dept., \\
Denys Wilkinson Building, \\ 
Keble Rd, Oxford, OX1 3RH, UK\\ 
E-mail: a.cooper-sarkar@physics.ox.ac.uk}

\maketitle

\abstracts{Use of HERA jet production cross-sections can extend our knowledge 
of the gluon parton distribution function and yield accurate measurements of 
$\alpha_s(M_Z)$ in addition to illustrating the running of $\alpha_s$ with scale.
}

\section{Introduction}

\begin{figure}[tbp]
\vspace*{13pt}
\begin{center}
\begin{tabular}{ll}
\psfig{figure=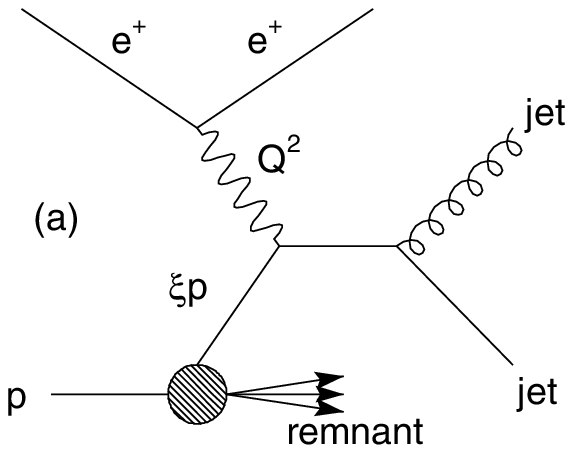,width=0.3\textwidth}~~~~ &
\psfig{figure=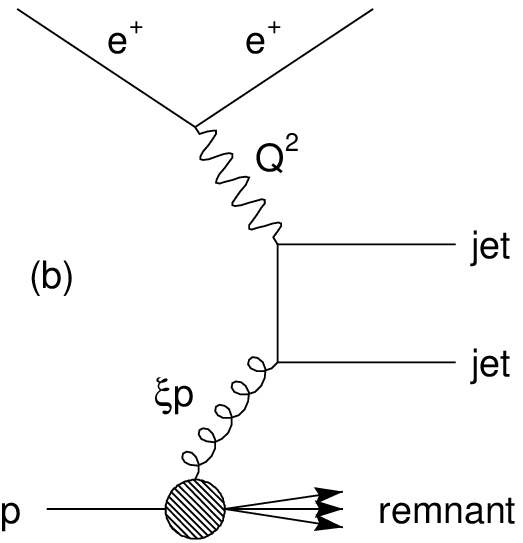,width=0.25\textwidth} \\
\end{tabular} 
\caption{Leading order QCD diagrams from dijet production in DIS.
(a) QCD Compton; (b) Boson-gluon fusion.}
\label{fig:lo-jets-diag}
\end{center}
\end{figure}

The QCD processes that give rise to scaling violations in DIS
at leading order, namely QCD Compton (QCDC) 
and boson--gluon fusion (BGF),
also give rise to distinct jets in the final state provided that
the energy and momentum transfer are large enough. These processes 
are illustrated in Fig.~\ref{fig:lo-jets-diag}.
Both BGF and QCDC depend on $\alpha_s(M_Z)$, but the BGF process also 
depends directly on the gluon parton distribution function (PDF), whereas
QCDC depends on quark PDFs. 
The QCDC process dominates at large $Q^2$, where the most
important quark PDFs are the well-determined $u$ and $d$ valence 
distributions (for HERA kinematics).
Thus measuring jet rates at large $Q^2$ allows a 
determination of $\alpha_s(M_Z)$,
with reduced uncertainty from the less well known gluon PDF. Such measurements
of $\alpha_s(M_Z)$ are described in Sec.~\ref{sec:alphas}. 

Conversely, if $\alpha_s(M_Z)$ is well known, then the BGF process gives direct 
information on 
the gluon PDF. Historically, PDF fits have been made using 
NLO QCD in the DGLAP formalism to predict the inclusive cross-sections for 
world DIS data, but the gluon PDF affects these cross sections only 
indirectly through the scaling violations. Inputting jet data to such 
fits can improve the determinations of the gluon distribution. HERA jet data 
have recently been used for this purpose in the
ZEUSJETS PDF fit, as described in Sec.~\ref{sec:PDFS}. Clearly the combination
of inclusive cross-section data and jet production data could lead to a 
simultaneous determination of $\alpha_s(M_Z)$ and the gluon PDF.
An early analysis was performed by H1~\cite{h1asjets}, and this idea 
has recently been 
fully developed into a simultaneous fit of $\alpha_s(M_Z)$ and all the PDFs in
the ZEUSJETS-$\alpha_s$ fit~\cite{zeusjets}, as also described in 
Sec.~\ref{sec:PDFS}.

\section{Measurements of $\alpha_s$ from jet production data}
\label{sec:alphas}

\begin{figure}[tbp]
\vspace{2cm}
\begin{minipage}{8cm}
\vspace{-1.5cm}\rotatebox{270}{\includegraphics[width=0.45\textwidth]{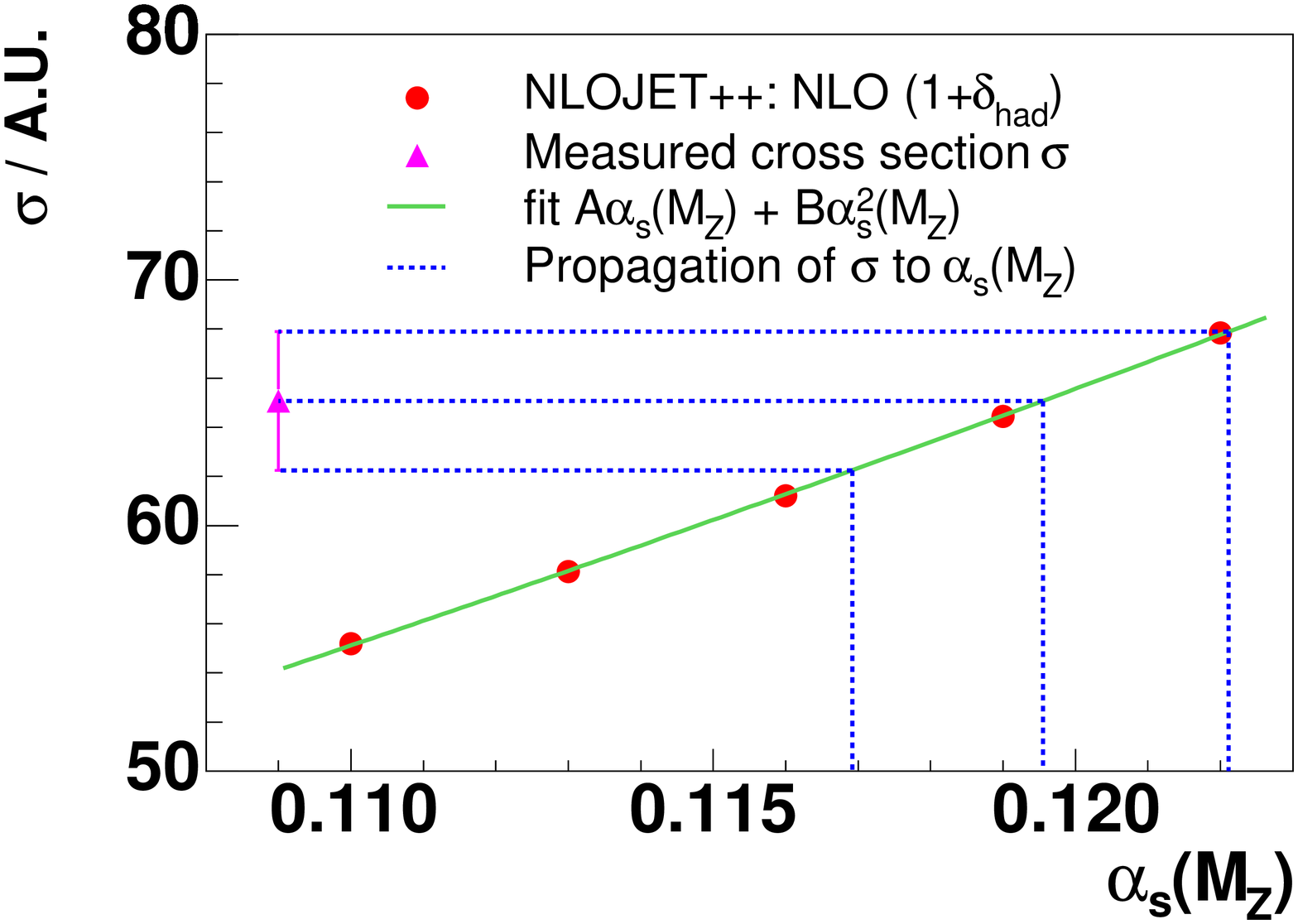}}
\end{minipage}
\hspace*{5.cm}\begin{minipage}{11cm}
\vspace{-5cm}{\includegraphics[width=0.65\textwidth]{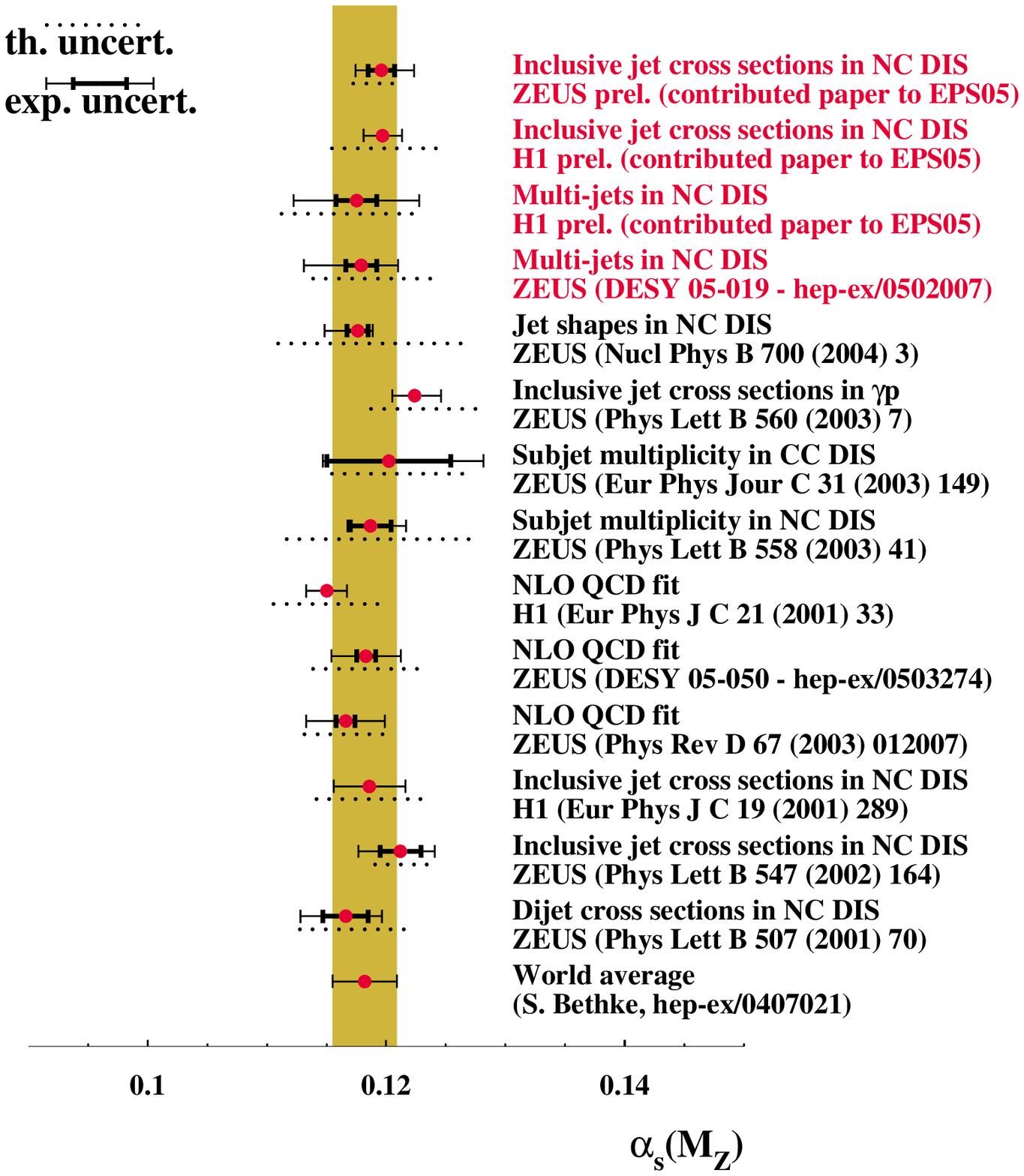}}
\end{minipage}
\caption{Left side: the method for determining $\alpha_s(M_Z)$. 
Right side: HERA $\alpha_s(M_Z)$ measurements, compared to the world average}
\label{fig:method/alf}
\end{figure}

The method used to extract $\alpha_s$ from jet measurements is nicely 
illustrated on the left hand side of Fig.~\ref{fig:method/alf}.
A programme, like DISENT or NLOJET, 
is used to calculate NLO QCD cross-sections 
for jet production, for several fixed values of $\alpha_s(M_Z)$. The parton 
distribution functions (PDFs) input to these calculations must be compatible 
with these $\alpha_s(M_Z)$ values. 
(In practice the CTEQ4A, MRST99 and ZEUS-S variable 
$\alpha_s$ series have been used).
The predictions for each cross-section bin, $i$, are fitted to the form
\begin{equation}
 \sigma_i(\alpha_s(M_Z)) = A_i \alpha_s(M_Z) + B_i \alpha_s^2(M_Z),
\end{equation} 
to determine $A_i$ and $B_i$ so that we have a prediction for any value of 
$\alpha_s(M_Z)$. We then map the measured value of $\sigma_i$ and its 
statistical error onto this function to determine $\alpha_s(M_Z)$ and its 
statistical error, as illustrated in the Figure. 
Experimental systematic uncertainties are evaluated by repeating the procedure
with the measurement shifted according to each source of systematic 
uncertainty and theoretical uncertainties are evaluated by repeating the 
procedure
with the predictions shifted according to different theoretical assumptions.

A summary of HERA 
$\alpha_s(M_Z)$ determinations, including those presented at EPS05,
is given on the right hand side of Fig.~\ref{fig:method/alf}.
A HERA average has been constructed from these results~\cite{claudia}, using 
the known experimental correlations within each experiment and assuming 
full correlation 
of theoretical scale uncertainties but no correlation between ZEUS 
and H1 data. The result is
\[ \alpha_s(M_Z) = 0.1186 \pm 0.0011 ({\rm exp.}) \pm 0.0050({\rm th.}) \]

The most accurate results come from inclusive jet production and 
tri-jet/di-jet/single jet ratios. The recent results for these 
processes are now discussed in more detail.

\begin{figure}[tbp]
\includegraphics[width=0.5\textwidth]{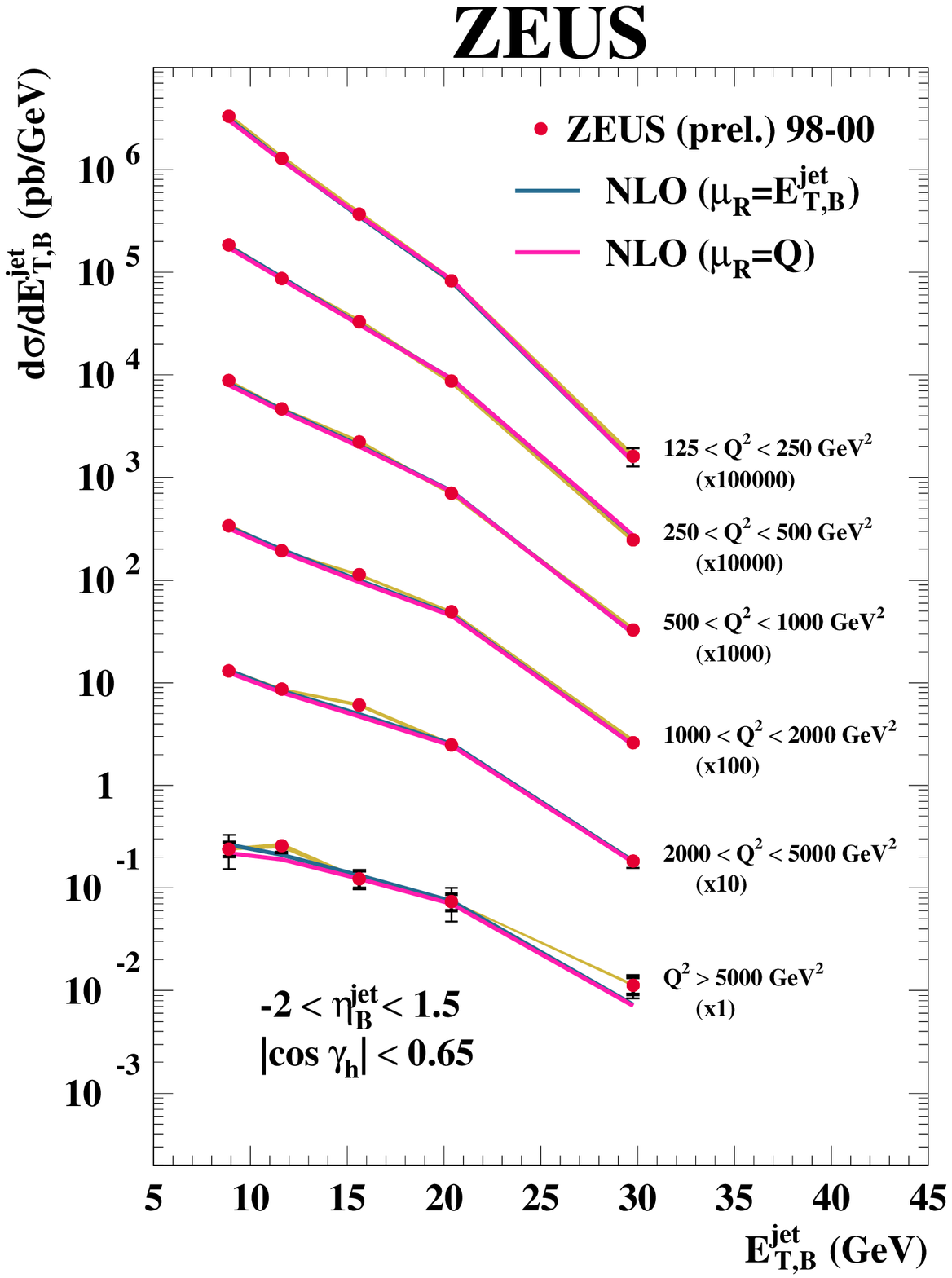}
\hspace*{5.5cm}\begin{minipage}{8cm}
\vspace{-6.5cm}
\rotatebox{270}{\includegraphics[width=0.5\textwidth]{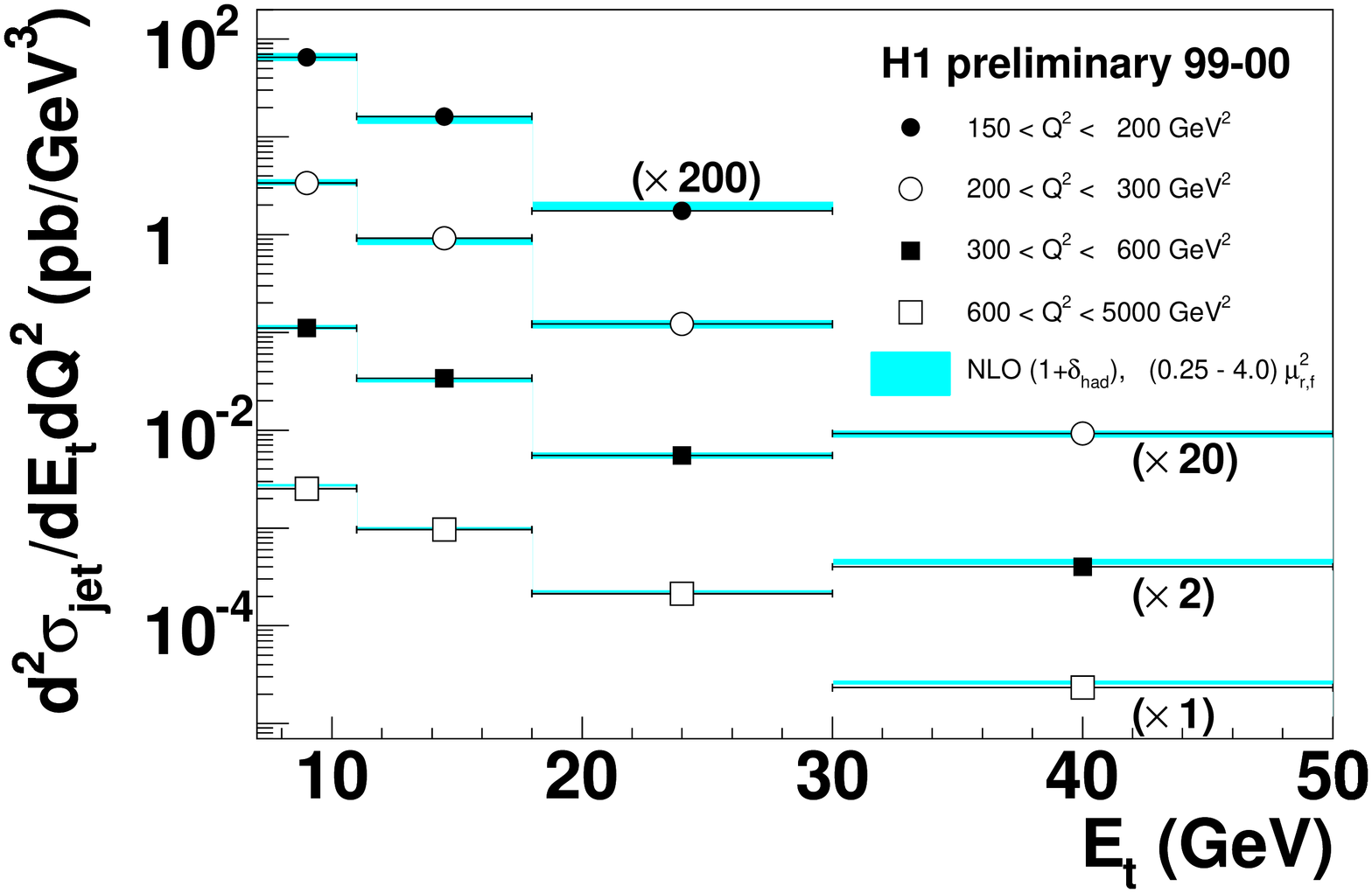} }
\end{minipage}
\vspace{-0.5cm}
\caption{$d\sigma/dE_{TB}$ for inclusive jet production. Left side: ZEUS data.
Right side: H1 data}
\label{fig:ddifdisjets}
\end{figure}

ZEUS have measured inclusive jet production in DIS using 
$81.7pb^{-1}$ of data from the 98-00 running period~\cite{9800incdis}. 
The cross-sections 
$d\sigma/dE^{jet}_{TB}$, where $E^{jet}_{TB}$ 
is the transverse energy of the jet in the Breit frame, 
are compared to the predictions of DISENT, 
for various $Q^2$ regions, in Fig.~\ref{fig:ddifdisjets}. The method outlined 
above has been used to extract $\alpha_s(M_Z)$. The best result is obtained for
$Q^2 > 500$ GeV$^2$, when the theoretical uncertainties are minimized:
\[ \alpha_s(M_Z) = 0.1196 \pm 0.0011 ({\rm stat.}) \pm^{0.0019}_{0.0025}({\rm sys.}) \pm^{0.0029}_{0.0017}({\rm th.}) \]
This can be compared to the previous ZEUS result from a similar treatment of 
data from the 96-97 running period:
\[ \alpha_s(M_Z) = 0.1212 \pm 0.0017 ({\rm stat.}) \pm^{0.0023}_{0.0031}({\rm sys.}) \pm^{0.0029}_{0.0027}({\rm th.}) \]
Not only do the higher statistics of the newer data reduce the statistical 
error, they also allow a reduced estimate of the systematic error. The biggest 
contribution to the experimental systematic error comes from the absolute 
energy scale of the jets. For ZEUS this is $\sim 1\%$, leading to a 
$\sim5\%$ error in the jet cross-sections. The theoretical uncertainties come 
from the usual procedure of varying the renormalisation and factorisation 
scales ($\mu_R = \mu_F = E_T$) by a factor of two. Fig.~\ref{fig:ddifdisjets} 
also illustrates similar data from
H1 from $61 pb^{-1}$ of 99/00 running~\cite{9900incdis}. 
The value of $\alpha_s(M_Z)$ extracted from these data is
\[ \alpha_s(M_Z) = 0.1175 \pm 0.0016 ({\rm exp.}) \pm^{0.0046}_{0.0048}({\rm th.}) \]
 
\begin{figure}[tbp]
{\includegraphics[width=0.5\textwidth]{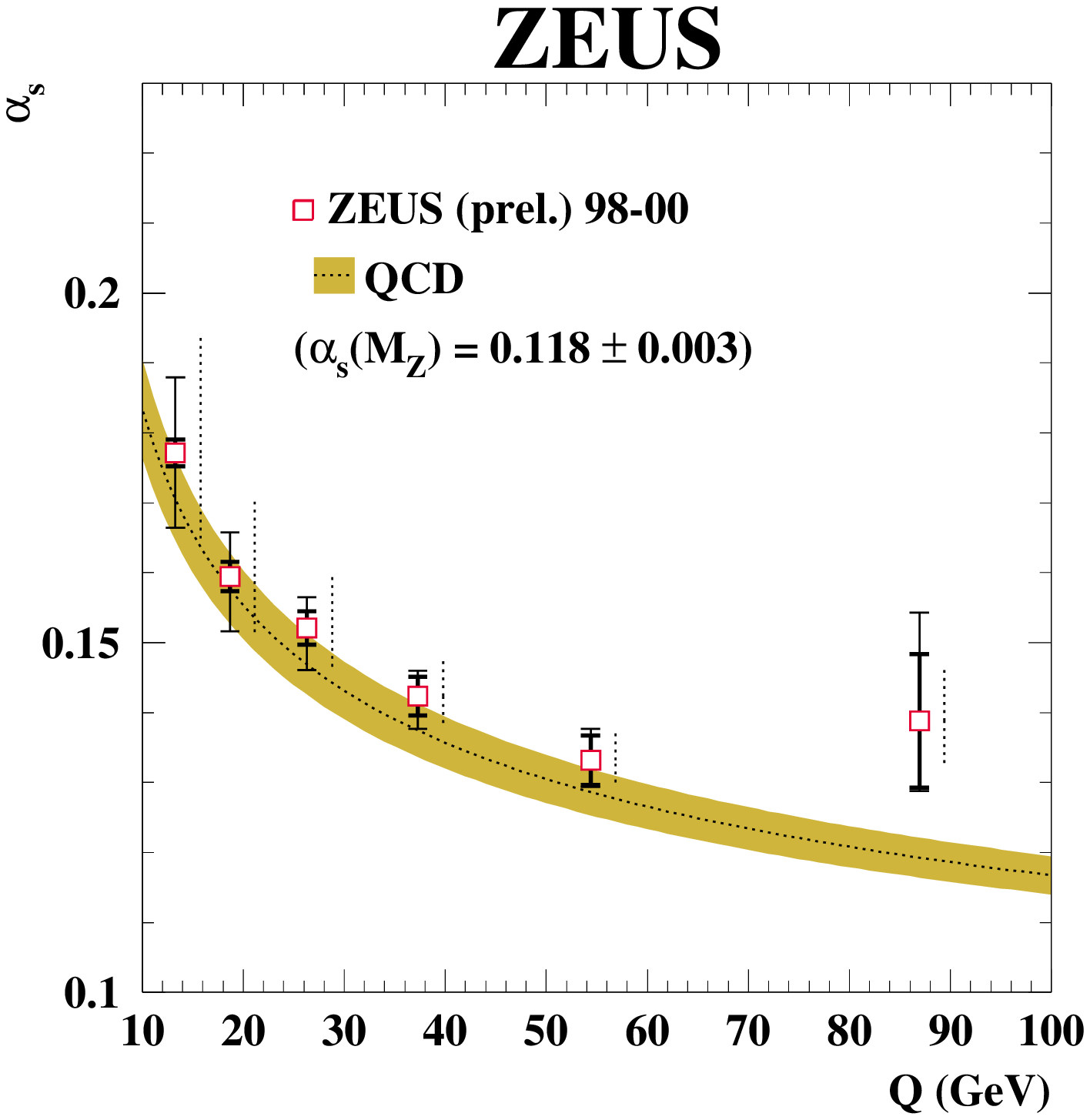}}
\hspace*{5.5cm}\begin{minipage}{8cm}
\vspace{-6.5cm}
\includegraphics[width=0.8\textwidth]{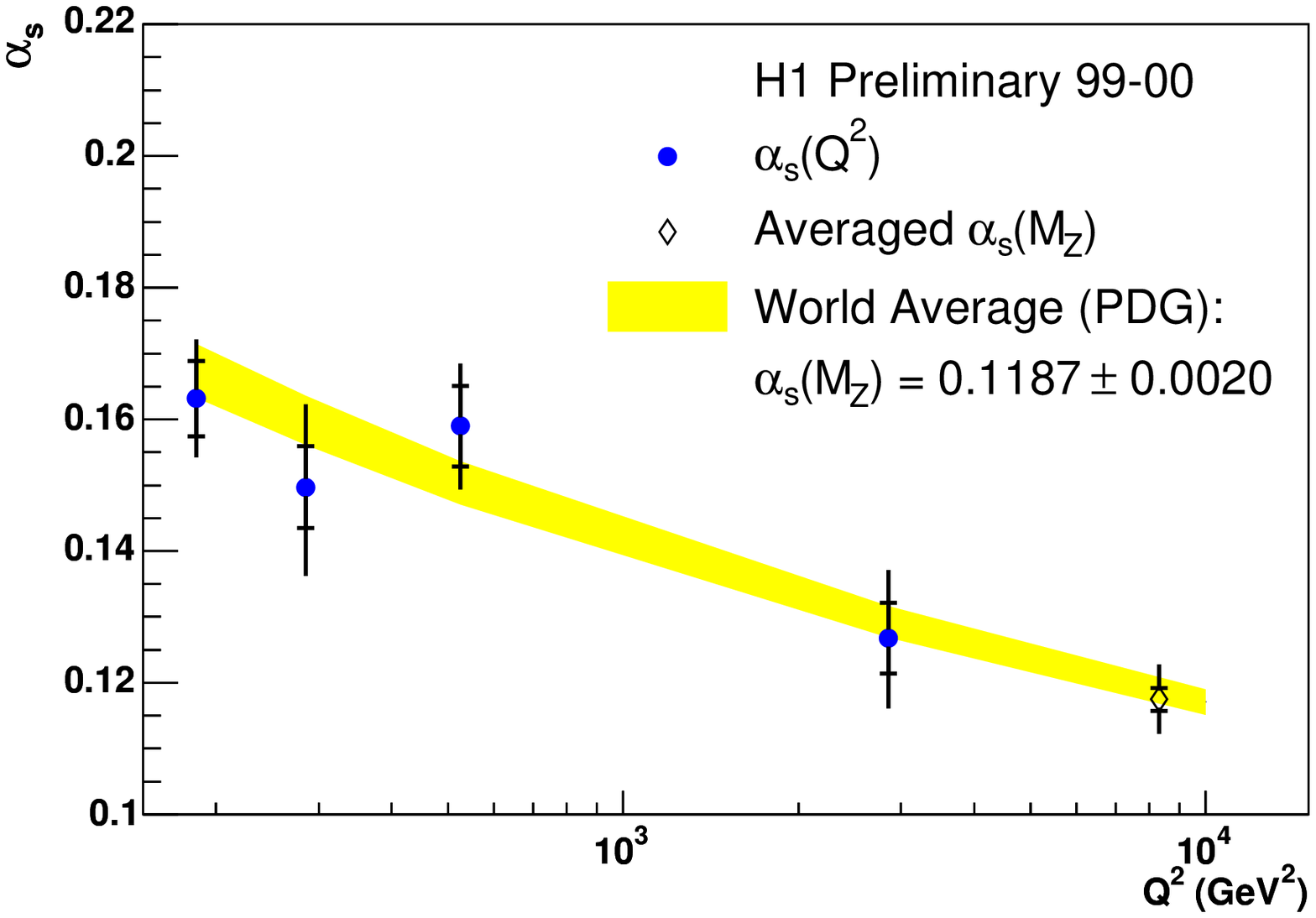}
\end{minipage}
\vspace{-1cm}
\caption{Left side: illustration of running $\alpha_s$ from ZEUS analysis of 
inclusive jet data. Right side: Illustration of running $\alpha_s$ from H1 
analysis of tri-jet to di-jet cross-section ratios }
\label{fig:alfscale}
\end{figure}

The method used to extract $\alpha_s$ can also be used to fit for 
$\alpha_s(<E_T>)$ or $\alpha_s(<Q^2)>$. Fig.~\ref{fig:alfscale} illustrates 
these results 
as a function of $Q^2$, demonstrating the running of $\alpha_s$ with scale, 
from data within a single experiment. 

Both experiments have used data from the same running periods to measure 
mutlijet production~\cite{zeusmj,h1mj}. 
Previously the ratio of di-jet to 
single-jet cross-sections has been used to extract $\alpha_s$. However, 
increased statistics now allows us to use tri-jet to di-jet cross-sections and 
these are more sensitive to $\alpha_s$. The advantage of the ratio technique 
is that many correlated experimental and theoretical uncertainties cancel out 
in ratio. The usual method to extract $\alpha_s$ is applied to predictions for
$R_3/R_2(\alpha_s(M_Z))$. The result for ZEUS is:
\[ \alpha_s(M_Z) = 0.1179 \pm 0.0013 ({\rm stat.}) \pm^{0.0028}_{0.0046}({\rm sys.}) \pm^{0.0054}_{0.0046}({\rm th.}) \]
and for H1:
\[ \alpha_s(M_Z) = 0.1175 \pm 0.0017 ({\rm stat.}) \pm 0.005({\rm sys.}) \pm^{0.0054}_{0.0058}({\rm th.}) \]
As usual the major contribution to the experimental systematic error is the 
jet energy scale 
uncertainty, and for the theoretical uncertainty it is the remormalisation and
factorisation scale uncertainty. Fig.~\ref{fig:alfscale} also 
illustrates the running of 
$\alpha_s$ from the H1 analysis of multijet production.

\section{Use of jet production data to determine the gluon and $\alpha_s(M_Z)$ in PDF fits}
\label{sec:PDFS}

Both HERA collaborations have recently made PDF fits 
using exclusively their own data~\cite{zeusjets,h103038}. The 
advantages of a fit within one experiment are that the systematic uncertainties
are well understood, and the advantages of a HERA only fit are that these 
data have the most extensive kinematic coverage of any DIS data, on a pure 
proton target, hence without the uncertainties of heavy target corrections, 
and without the need to make strong isospin assumptions. 
However, such HERA only analyses have only been possible since the 98/00 
running extended HERA kinematic coverage to high $Q^2$, where the contributions
 of $W^{\pm}$ and $Z$ exchange become important. This allows us to gain 
information on the valence $d$ and $u$ quarks at high $x$, from the $e^+p$ 
and $e^-p$ charged current cross-sections, respectively. Furthermore, 
the difference between the $e^+p$ and $e^-p$
high-$Q^2$ neutral current cross-sections allows us
to extract the valence structure function $xF_3$ for all $x$.
Valence information was the missing piece in a HERA only analysis, since HERA 
data have provided the major contributions to the measurement of the low-$x$ 
sea and gluon distributions, through the structure function, $F_2$, and its
scaling violations, $dF_2/dln(Q^2)$, respectively. 
It has been a limitation of all PDF fits, 
including the MRST and CTEQ global fits to world data, 
that information on the high-$x$ gluon is lacking.
The global fits have used Tevatron jet data to remedy this.
However, the HERA jet data have the advantage of 
much smaller experimental uncertainties 
and do not suffer from possible uncertainties due to new physics.

The ZEUSJETS PDF fit~\cite{zeusjets} 
uses all the inclusive double differential
cross-section data from HERA-I running (94-00) and adds the DIS inclusive jet
production data from~\cite{disjetsold} and the di-jet data 
from photoproduction~\cite{gammadijets} from 96-97 running, 
to gain information on the gluon. As 
remarked earlier, the jet cross-sections depend on the gluon and quark PDFs 
and on $\alpha_s$, and this can be used to break the strong coupling between
$\alpha_s$ and the gluon PDF which plagues the indirect extractions from the
scaling violations of inclusive data. The fit takes full account of the 
correlated systematic uncertainties within and between the data sets and these
are propagated into the PDF uncertainties using the OFFSET method.

A remark is in order on the use of photoproduced di-jet data. It is well known
 that at leading order photoproduction can proceed via direct and resolved 
processes. In the former, the photon interacts directly as a point-like 
particle, whereas,
in the latter, the photon has decomposed into quarks and it is one of these
quarks which initiates the interaction. The resolved cross-sections will have 
some sensitivity to the photon PDF. We avoid this by restricting the analysis 
to direct enriched cross-sections by the cut 
\[x_\gamma^{obs} = \Sigma_i E_T^{jet_i} exp(-\eta^{jet_i})/ 2yE_e > 0.75 \] 
Of course at NLO these distinctions cannot be made precisely, but this cut 
still serves to minimise the influence of the choice of photon PDF.

A new technique had to be developed to include NLO jet cross-sections in a PDF
fit, since the computation of such cross-sections is very CPU intensive. 
Hence the NLO programmes were used to compute LO and NLO weights, 
$\tilde{\sigma}$, which are independent of $\alpha_s$ and the 
PDFs, and are obtained by integrating the corresponding partonic hard cross 
sections in bins of $\xi$ (the proton momentum fraction carried by
the incoming parton), $\mu_F$ and $\mu_R$.
The NLO QCD cross sections, for each measured bin, were then 
obtained by folding these weights with the PDFs and $\alpha_s$ according to the
formula
\begin{equation}
         \sigma = \sum_n \sum_a \sum_{i,j,k} 
                  f_a({\langle \xi \rangle}_i , {\langle \mu_F \rangle}_j)  
                  \cdot \alpha_s^n({\langle \mu_R \rangle}_k) 
                  \cdot \tilde{\sigma}^{(n)}_{a,\{i,j,k\}} ~,
\end{equation}
where the three sums run over the order $n$ in $\alpha_s$, the flavour $a$ of 
the incoming parton, and the indices ($i,j,k$) of the $\xi$, $\mu_F$ and 
$\mu_R$ bins, respectively. 
The PDF, $f_a$, and $\alpha_s$ were evaluated at the mean values 
${\langle \xi \rangle}$, ${\langle \mu_F \rangle}$ and
${\langle \mu_R \rangle}$ of the variables $\xi$,~$\mu_F$ and $\mu_R$ in each 
($i,j,k$) bin. This procedure reproduces the NLO predictions
 to better than $0.5\%$.

\begin{figure}[tbp]
{\includegraphics[width=0.45\textwidth]{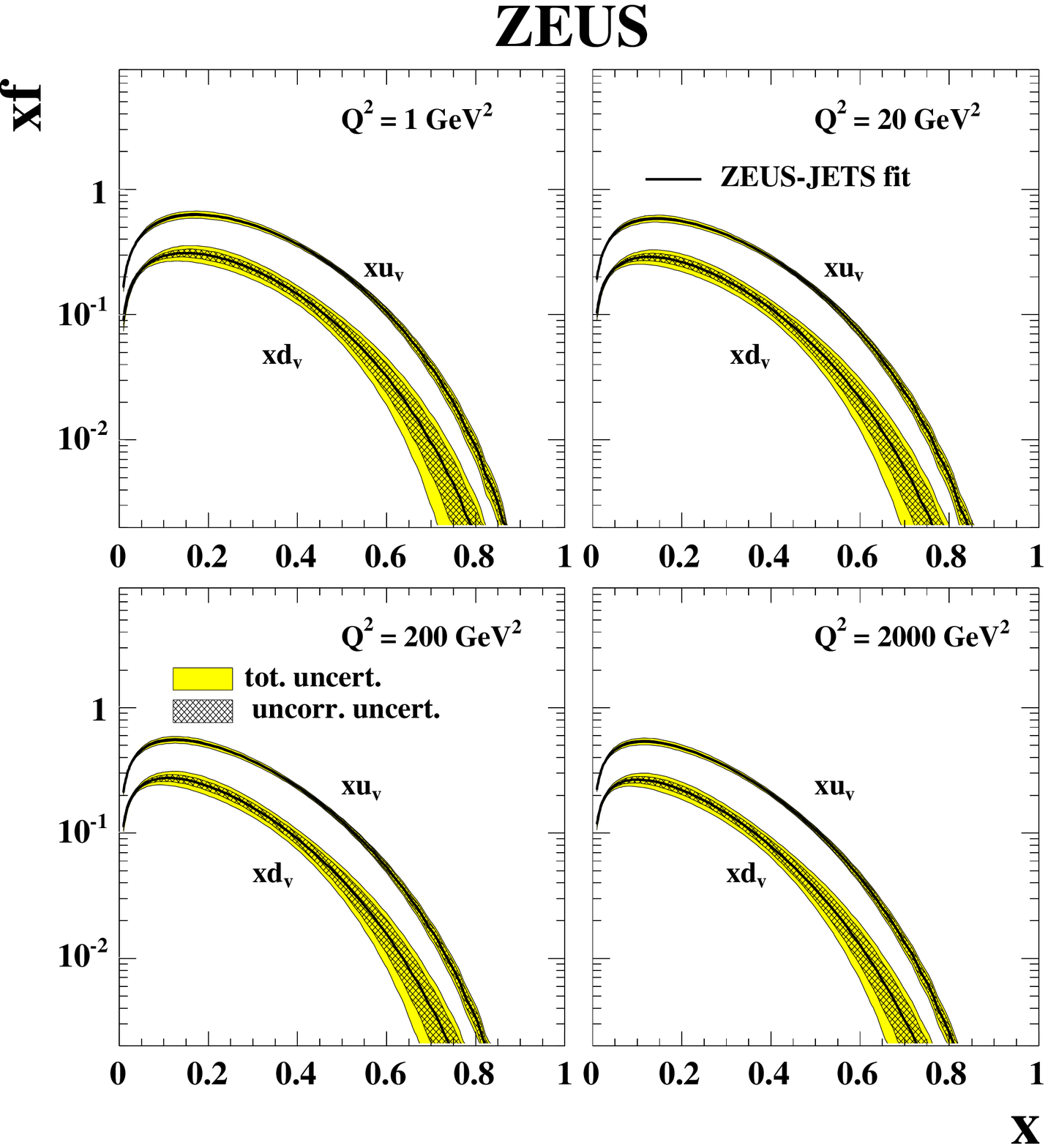}}
\hspace*{6cm}\begin{minipage}{11cm}
\vspace{-7cm}{\includegraphics[width=0.5\textwidth]{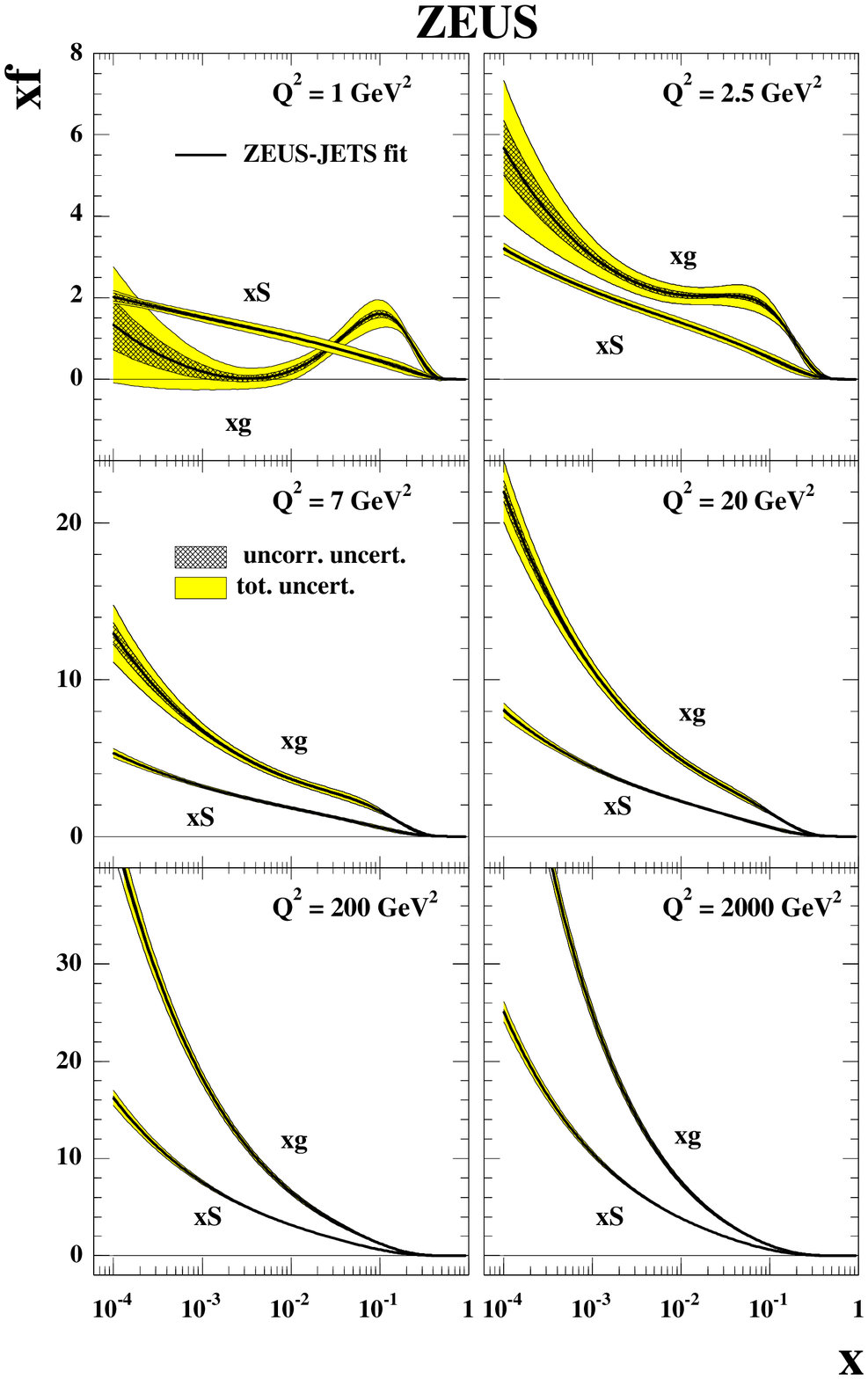}}
\end{minipage}
\caption{PDFs from the ZEUSJETS PDF fit. Left side: $u$ and $d$ valence PDFs. Right side: sea and gluon PDFs }
\label{fig:valseaglu}
\end{figure}

\begin{figure}[Ht]
\centerline{\psfig{figure=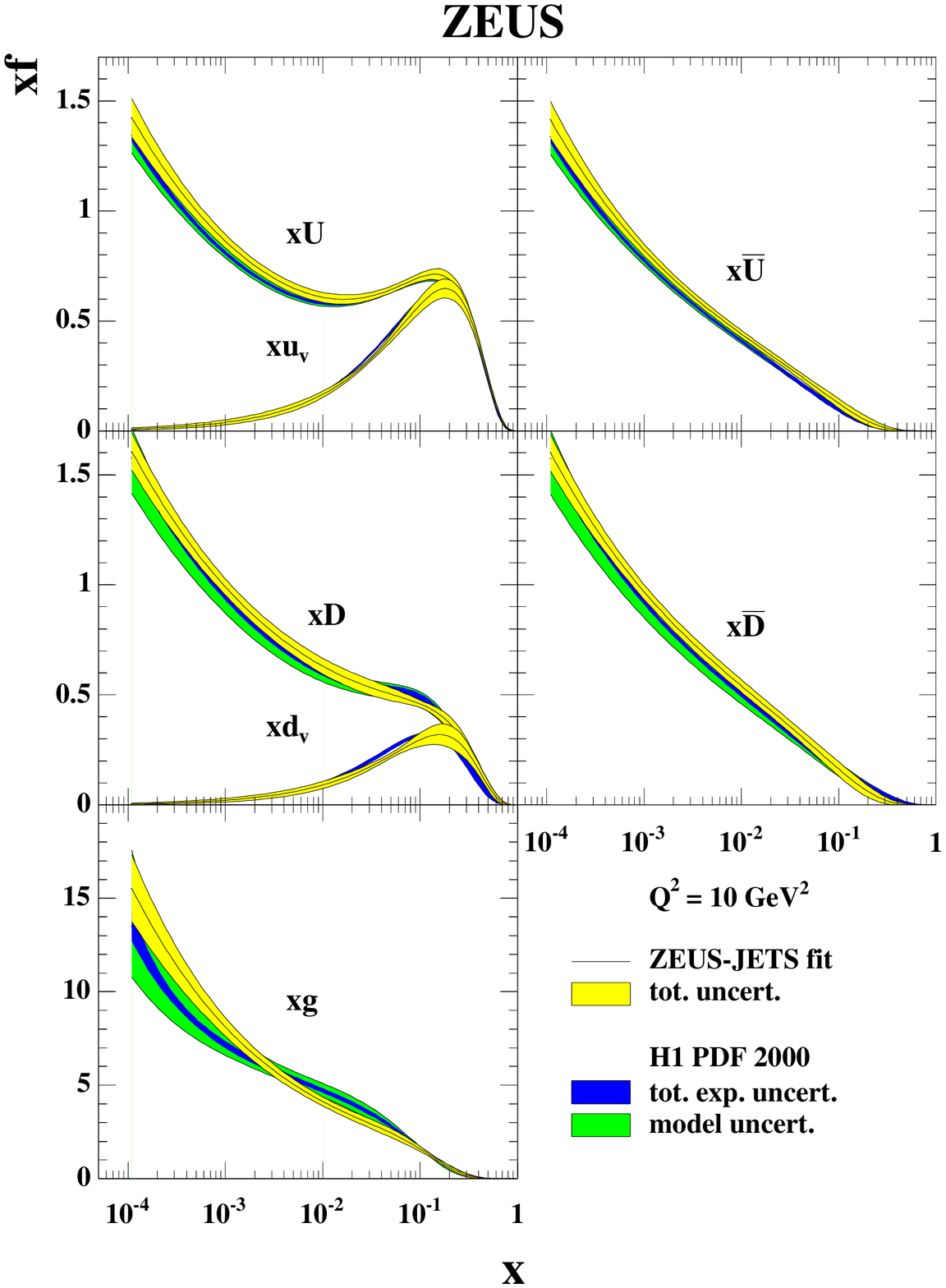,width=0.49\textwidth}
\psfig{figure=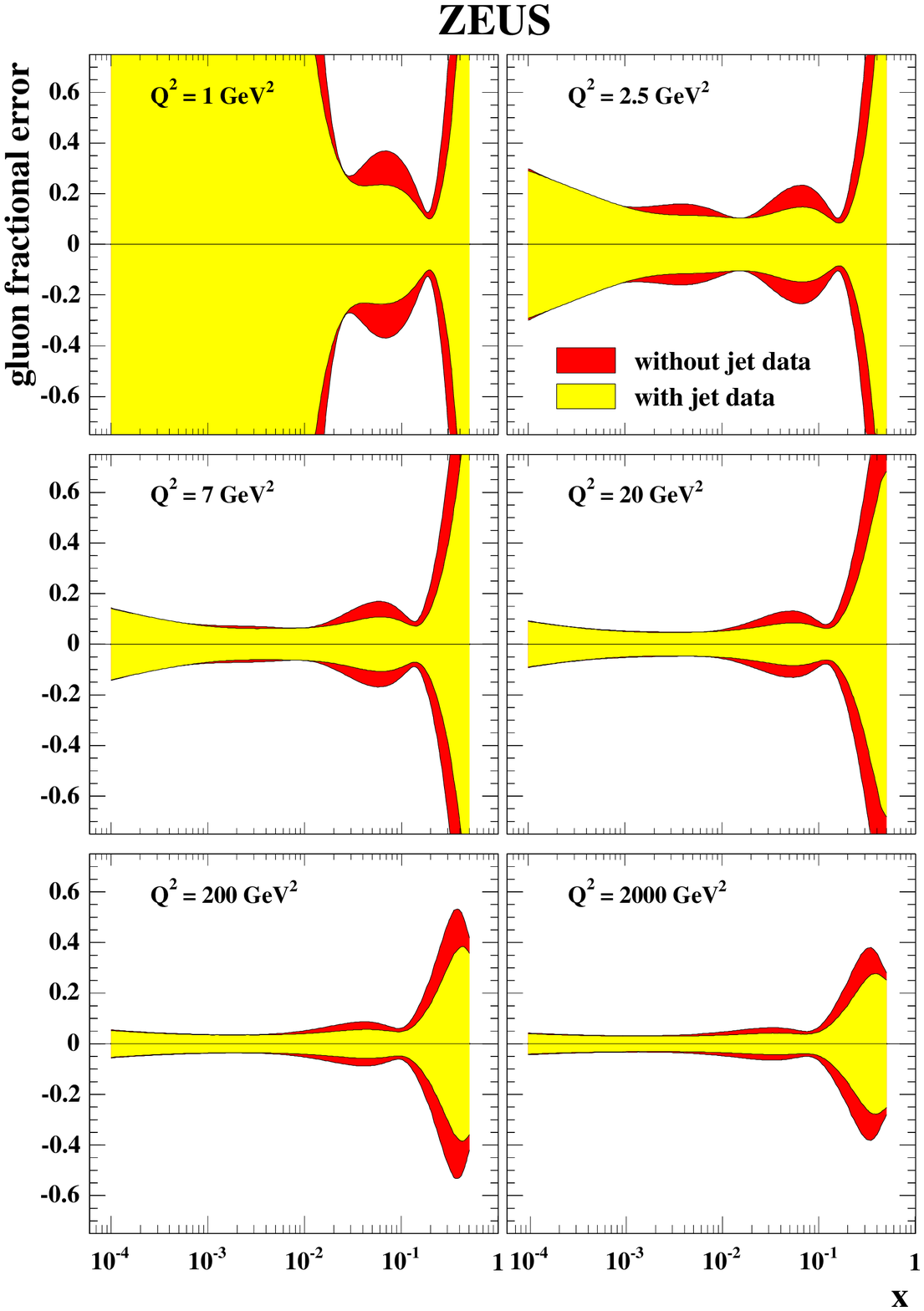,width=0.45\textwidth} }
\caption{Left side: ZEUSJETS PDFs compared to H1 PDFs 2000. Right side: 
the fractional uncertainties on the gluon PDF are shown before and after the 
jet data are included in the ZEUSJETS PDF fit.}
\label{fig:gluimph1}
\end{figure}

The ZEUSJETS PDF fit gives a very good $\chi^2$ to all of the input data sets,
simultaneously describing inclusive cross-sections and jet production data,
thus providing a compelling demonstration of QCD factorisation. 
The extracted PDFs are shown for various $Q^2$ in Fig.~\ref{fig:valseaglu}.
The ZEUSJETS PDFs are in good agreement with world PDF extractions and are 
shown compared to the H1 PDFs in Fig.~\ref{fig:gluimph1}. This right hand side
of this figure illustrates how the input of the jet data has improved 
the uncertainty on the gluon PDF, by about a factor of two, 
in the region $0.1 < x < 0.4$.
\begin{figure}[htb]
\begin{center}
\begin{tabular}{ll}
\psfig{figure=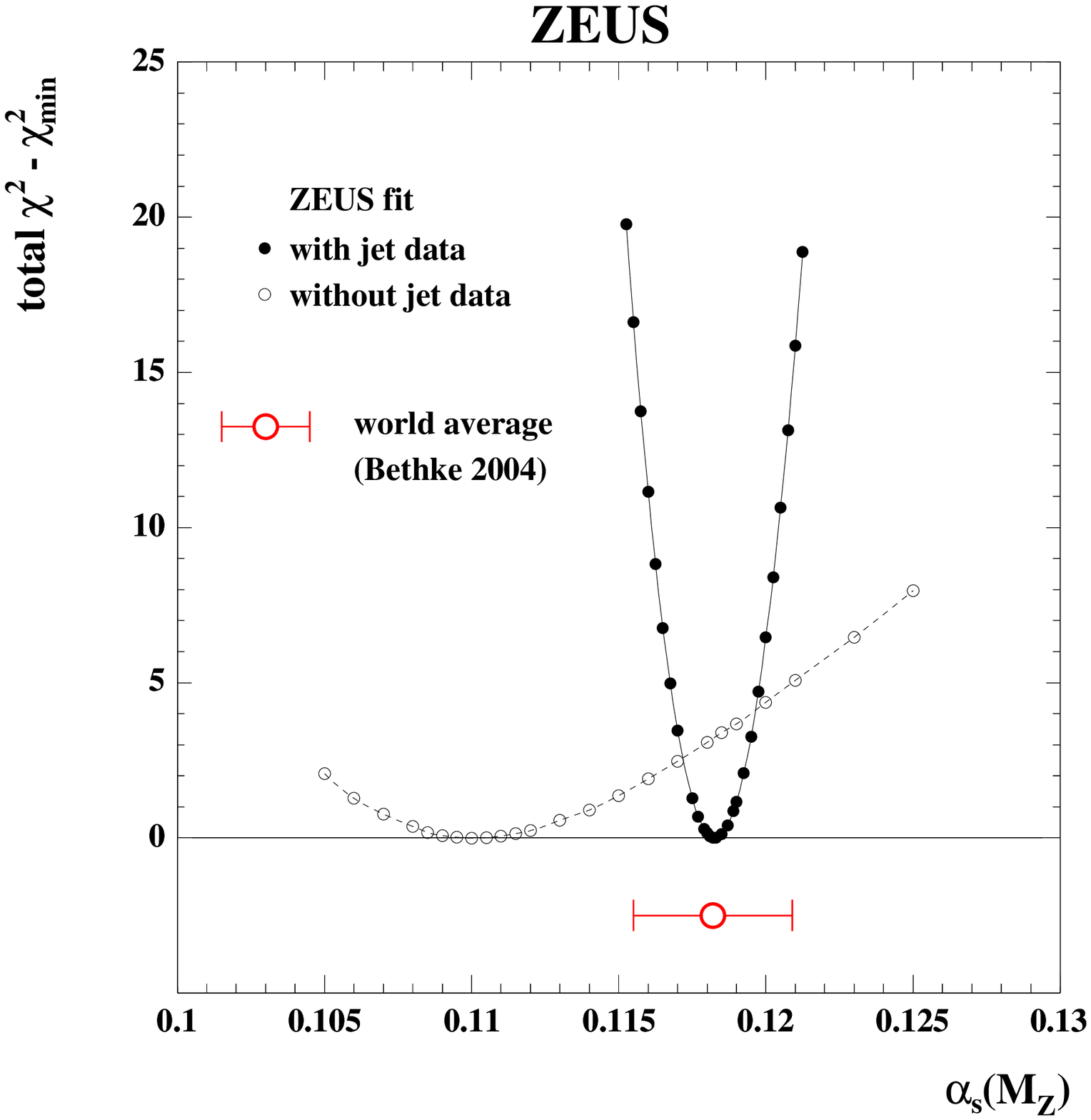,width=0.5\textwidth}~ &
\psfig{figure=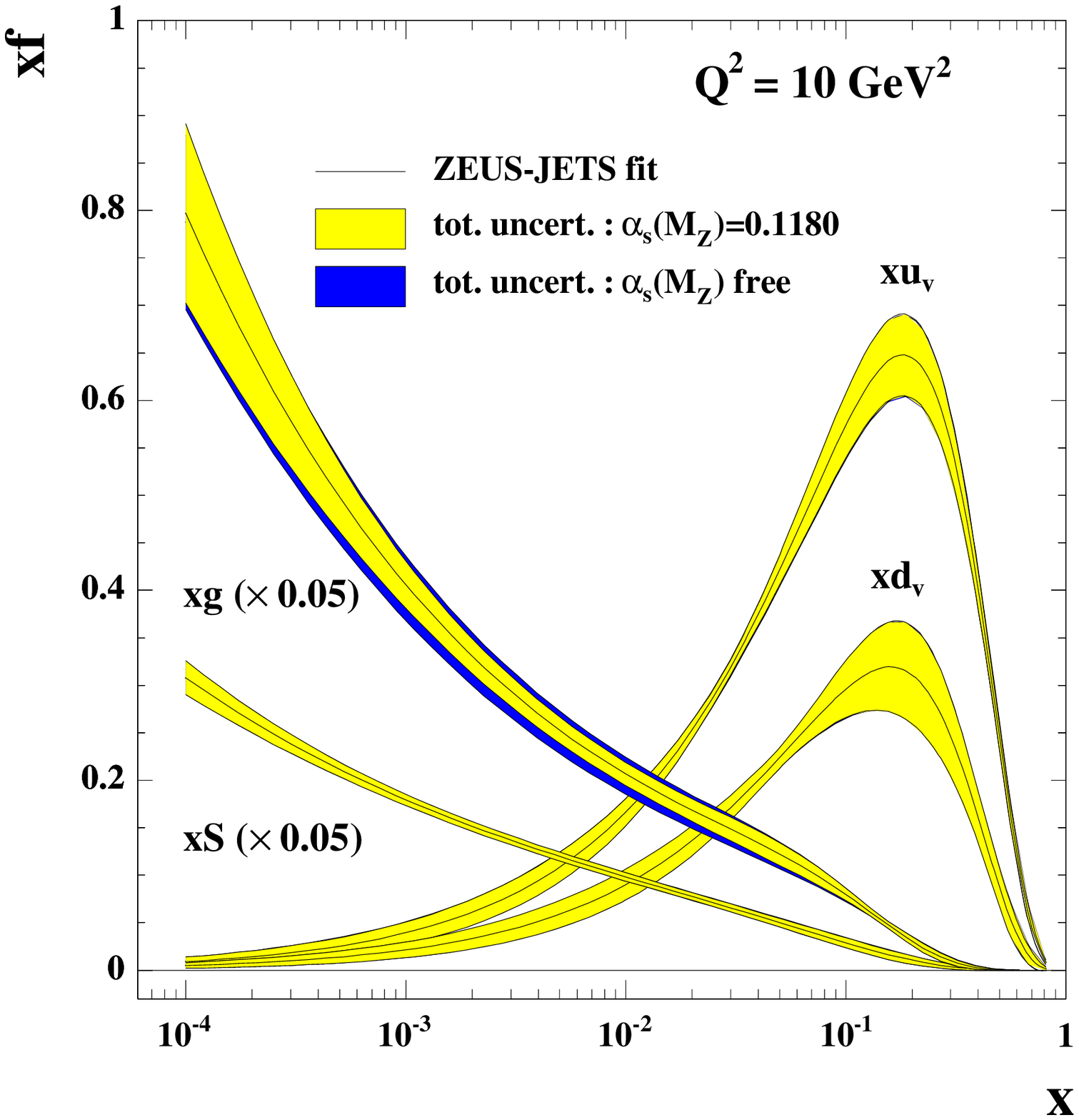,width=0.5\textwidth} \\
\end{tabular} 
\caption{Left side: $\chi^2$ profile for $\alpha_s(M_Z)$ with and without jet 
data. Right side: the ZEUS JETS$-\alpha_s$ PDFs with the uncertainty from 
variation of $\alpha_s(M_Z)$ included. }
\label{fig:alfchipdf}
\end{center}
\end{figure}

\begin{figure}[tbp]
\begin{center}
\begin{tabular}{ll}
\psfig{figure=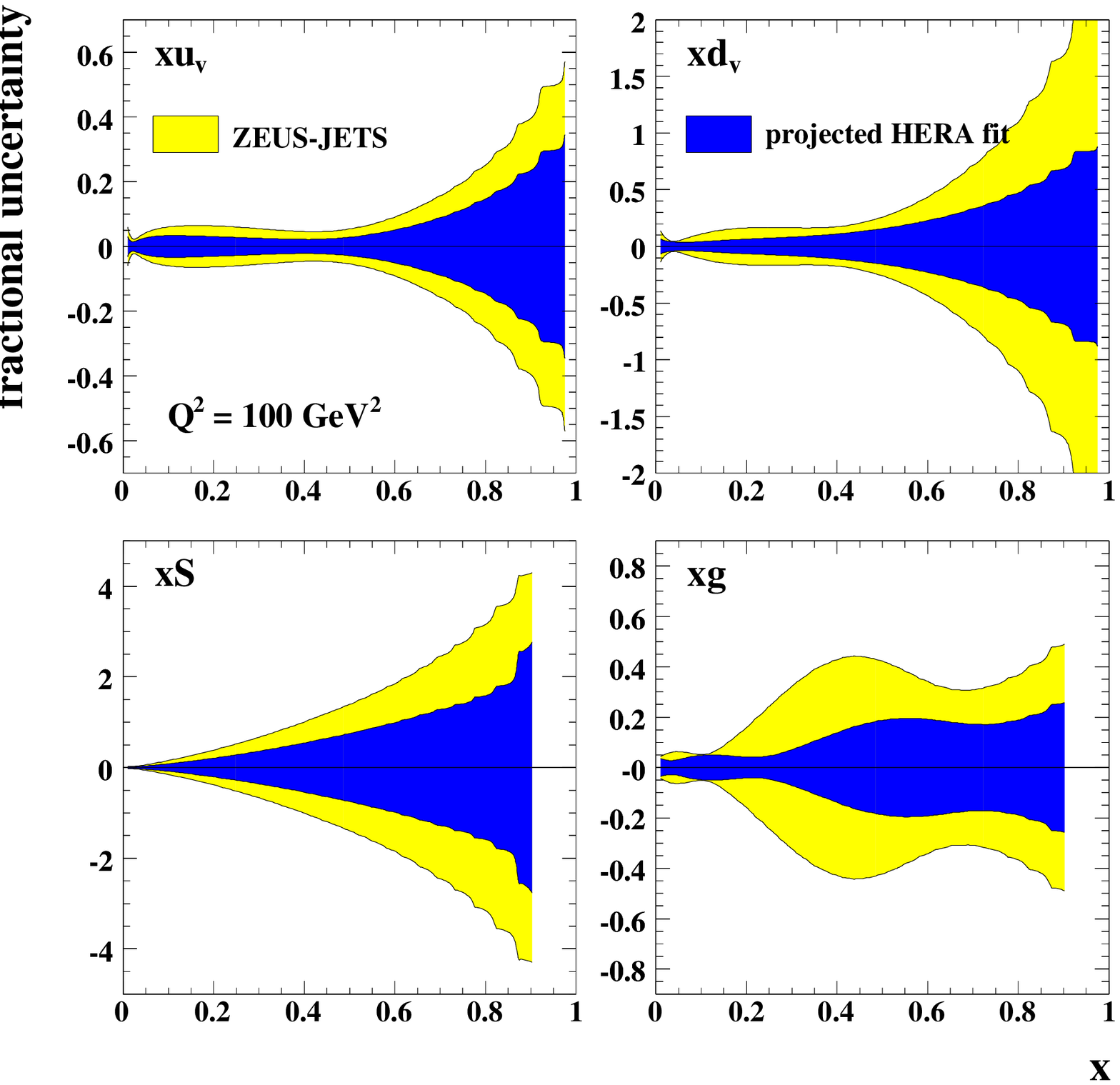,width=0.7\textwidth}
\end{tabular} 
\caption{Projected PDF improvements after HERA-II running. }
\label{fig:improve}
\end{center}
\end{figure}
Most world PDF fits are performed assuming a fixed value of $\alpha_s(M_Z)$ 
equal to the world average ($\alpha_s(M_Z) = 0.118$ for the ZEUSJETS fit 
described above). This is because the coupling between the gluon density and 
the value of $\alpha_s(M_Z)$ in the DGLAP equations does not allow a very 
accurate measurement of $\alpha_s(M_Z)$. 
However, with the added information on the gluon which comes from the jet data,
 we can make an accurate extraction by allowing $\alpha_s(M_Z)$ to be a free
parameter simultaneously with all the PDF parameters, in the 
ZEUSJETS-$\alpha_s$ fit:
\[ \alpha_s(M_Z) = 0.1183 \pm 0.0007 ({\rm stat.}) \pm 0.0027({\rm sys.}) \pm 0.0008({\rm model}) \pm 0.005({\rm th.}) \]
where the systematic error contains all the correlated experimental systematics of the contributing data sets. The model error includes uncertainties due to
changes in the choice of the proton PDF parametrisation at the starting 
scale, and in the choice of the photon PDF. 
The theoretical error comes from variation 
of the choice of the renormalisation and factorisation scales as usual.
Fig.~\ref{fig:alfchipdf} illustrates the contribution of the jet data to the 
accuracy of the $\alpha_s$ extraction by comparing the $\chi^2$ profile, with 
and without the jet data. This is the first extraction from HERA data alone
because the extractions described in Sec.~\ref{sec:alphas} 
all require input of 
PDFs which were extracted using other data. In the ZEUSJETS fit we are able to
vary the $\alpha_s(M_Z)$ used for predicting the jet cross-sections, 
simultaneously with varying the PDFs. A consequence of this 
is that the uncertainties on the PDFs due to the uncertainty on 
$\alpha_s(M_Z)$ is much reduced in the ZEUSJETS-$\alpha_s$ fit, as
also illustrated in Fig.~\ref{fig:alfchipdf}.

\section{Outlook}

We can look forward to improving these PDF fits with more jet data in the near 
future, since the jet data included in the present fit came only from the 
96-97 running. Jet data from the 98-00 running are already becoming available,
as described in Sec.~\ref{sec:alphas}, where they were used for 
preliminary $\alpha_s$ extractions~\cite{9800incdis,9900incdis}. Combination 
of ZEUS and H1 data will also yield greater accuracy.
In the mid-term future we look forward to improving these measurements with 
HERA-II data. A study of possible improvements~\cite{claire}
 has assumed the achievable goals of a 
luminosity of $350 pb^{-1}$ for both $e^+$ and $e^-$ running, 
and $500 pb^1$ of jet data. The extra statistics of high $Q^2$ data will yield 
improved high-$x$ valence information, and an extraction of $xF_3$ will 
extend the valence information down to $x \sim 0.01$. The extra statistics 
will also improve the high-$x$ sea extraction, whereas the jet data will 
further 
improve the high-$x$ gluon.  Fig.~\ref{fig:improve} shows the expected 
level of improvement in PDF uncertainty.

\end{document}